\documentclass[11pt,fleqn]{article}

\usepackage{psfrag}
\usepackage{graphicx}
\usepackage{amssymb}

\def\thefootnote{\fnsymbol{footnote}}

\newcommand{\be}{\begin{equation}}
\newcommand{\ee}{\end{equation}}
\newcommand{\bea}{\begin{eqnarray}}
\newcommand{\eea}{\end{eqnarray}}
\def\la{\langle}
\def\ra{\rangle}
\newcommand\p{\partial}
\newcommand\bp{\bar\partial}
\newcommand\NP[1]{:\! #1 \!:}
\newcommand\egal{&\!\!\!=\!\!\!&}
\renewcommand\t{\theta}
\renewcommand\tt{\tilde\theta}

\def\D{\Delta}
\def\I{\mathbb I}
\def\Z{\mathbb Z}
\def\R{\mathbb R}
\def\T{\mathcal{T}}
\def\V{\mathcal{V}}
\def\N{\mathcal{N}}
\def\L{\mathcal{L}}
\def\C{\mathcal{C}}
\def\K{\mathcal{K}}
\def\La{\Lambda}

\begin{document}

\rightline{}
\vskip 0.5cm
{\LARGE \centerline{Boundary height fields in the Abelian sandpile model}}
\vskip 1.5cm

\centerline{\large Geoffroy Piroux\footnote{piroux@fyma.ucl.ac.be} and Philippe
Ruelle\footnote{Chercheur Qualifi\'e FNRS; ruelle@fyma.ucl.ac.be}}

\vskip 0.5truecm
\centerline{Institut de Physique Th\'eorique}
\centerline{Universit\'e Catholique de Louvain}
\centerline{B--1348 \hskip 0.5truecm Louvain-La-Neuve, Belgium}

\vskip 2truecm
\abstract{We study the abelian sandpile model on the upper half plane, and
reconsider the correlations of the four height variables lying on the boundary. For more
convenience, we carry out the analysis in the dissipative (massive) extension of the model
and identify the boundary scaling fields corresponding to the four heights. We find that
they all can be accounted for by the massive pertubation of a $c=-2$ logarithmic
conformal field theory.}


\renewcommand{\thefootnote}{\arabic{footnote}}
\setcounter{footnote}{0}
\section{Introduction}

The description of equilibrium critical phenomena has been one of the greatest
success of two--dimensional conformal theories in the past twenty years \cite{cardy1}.  
More recently, attention has focused on new types of observables in otherwise well--known
models, and also on new kinds of critical systems. In both cases, non local features often
play an important role, because either the observables one is interested in are themselves
non local in terms of the natural microscopic variables, or else because the statistical
model possesses intrinsic non local properties. These studies often lead to a
description in terms of conformal theories with peculiar properties. A class of systems
with such properties is provided by sandpile models. Some of them, and in particular the
one we consider here, are believed to have a faithful description in terms of
logarithmic conformal field theories. 

Our motivation to study these models is two--fold. First, one wishes to see to what
extent they lend themselves to a conformal field theoretic approach, and if the adequacy
of the conformal description is as good as for the equilibrium systems. Second, the
logarithmic theories have been developed for themselves, but are complex and some of
their aspects are not fully understood yet. It should therefore be profitable to have 
concrete realizations in order to have a better understanding of the most
peculiar features. 

The sandpile model we consider here is the isotropic Abelian sandpile model (ASM), as
originally defined in \cite{btw}. The most natural variables to consider in a conformal
context are the four height variables. In the bulk, correlations of height 1 variables
can be handled by local calculations \cite{md1}, but height 2, 3 and 4 variables are much
more complicated, and only their one--site probabilities are known \cite{priez}. 
For sites on a boundary, closed or open, Ivashkevich showed, by using suitable
identities, that the non local configurations needed to handle the heights bigger than 1
could be reduced to local computations \cite{iv}. He was then able to compute the     
two--site joint probabilities of all height variables. He found that all correlations
decay like $r^{-4}$, and infered, wrongly as we will see, that all boundary height
variables scale to the same conformal field. 

Our aim in this paper is to revisit this problem and to identify the fields corresponding
to the boundary height variables, but in a dissipative extension of the model, known to be
described by a massive perturbation of the $c=-2$ conformal theory \cite{mr}. The main
advantage for doing this is that it allows for an unambiguous identification of the fields
from a few 2--point correlators, off, and therefore also at, criticality. The so--obtained
identification can then be checked from other 2--point and from 3--point functions. In
contrast, when one considers the critical, non dissipative model, the 2--point functions
yield ambiguous field identifications, which can only be fixed by using 3--point functions,
 and then checked from higher correlators. 

The article is organized as follows. The next section defines the model and sets our
conventions. Section 3 deals with the boundary unit height variable and what we call
supercritical height variables. They are much easier than the other ones, and illustrate
the way the identification with concrete fields is obtained. The identification of the
height 1 and supercritical height fields also facilitates that of the other heights.

Sections 4 and 5 form the hardcore of the article. In Section 4, we explain our
prescription ---a two--step burning algorithm--- to associate recurrent configurations of
the sandpile to spanning trees, from which a clear characterization of the local height
constraints follows.
We use it to compute 2--site probabilities for having a height 1 or a supercritical
height at one site, and any other height at the other site, from which we deduce the field
identification of all height variables. Section 5 checks these results by computing
explicitely all 2--site height probabilities in the massive extension of the sandpile
model, and certain 3--site probabilities involving a height 1 or a supercritical height. 

Since the bulk of the calculations reported here was done, an article by Jeng
\cite{jeng1} has appeared, where precisely the same problem is addressed.
The two works were carried out independently and differ in two ways. The first one is that
we study the dissipative model, while Jeng considers it at criticality. This fact
enables us to deduce the field identifications for the boundary height variables $h>1$
from the spanning tree characterizations of a single insertion of such variables,
technically much simpler to solve than the 2--site insertions. The latter are only used
as cross--checks, in contrast to the approach at criticality which needs them as inputs.
So working off criticality offers a simpler and more reliable access to the fields.
Secondly, we use a different characterization of the height variables bigger or
equal to 2 in terms of spanning trees, which is based on a modified, two--step burning
algorithm. This, we believe, leads to a more transparent formalism which generalizes to
multisite probabilities. Our results and conclusions however fully agree with those of Jeng.


\section{The dissipative sandpile model}

Consider a finite portion $\L$ of a square lattice and define at each site $i$ a
(sand) height variable $h_{i}$ which can take the integer values $1,2,3,\ldots$. A
configuration $\C$ of the sandpile is the set of height values $\{h_{i}\}$ for all
sites. The dynamics is defined in terms of a symmetric toppling matrix $\D$. Its entries
are all integers, positive on the diagonal, negative off the diagonal, and it has row
sums which are non negative. A configuration is called stable if all heights satisfy
$h_{i}\le \D_{ii}$. 

The system evolves in discrete time as follows. To the stable configuration $\C_{t}$ at
time $t$ we add a sand grain at a random site $i$ (chosen with uniform
distribution say), namely we set $h_{i} \to h_{i}+1$. This new configuration, if stable,
defines $\C_{t+1}$. If it is not stable, the unstable site $i$ topples: it loses
$\D_{ii}$ grains, every other site $j$ receives $-\D_{ij}$ grains, whereas  $\sum_{j\in
\L}\D_{ij}$ sand grains are dissipated (they fall off the pile, to a sink). That is, when a
site $i$ topples, we update the heights according to 
\begin{equation}
h_{j} \longrightarrow h_{j} - \D_{ij}\,, \qquad \forall j \in \L.
\end{equation}
If other sites become unstable after the toppling of the site $i$, they topple following
the same rule. All unstable sites are then toppled until the configuration becomes
stable again. This configuration is then taken as $\C_{t+1}$. In this way, the toppling
at the seeded site can trigger a potentially large avalanche, resulting in a
configuration $\C_{t+1}$ which can be completely different from $\C_{t}$.

Provided there are dissipative sites, {\it i.e.} sites $k$ for which $\sum_{j\in\L} 
\D_{kj}>0$, the dynamics is well defined: it does not depend on the order
in which the sites are toppled (the model is Abelian), and the relaxation of the seeded
configuration to $\C_{t+1}$ requires a finite number of topplings. 

One is generally interested in the long time behaviour of the sandpile. As shown by Dhar
\cite{dhar}, this behaviour is characterized by a unique time invariant probability
measure $P^{*}_{\L}$, which specifies the probabilities of occurrence of all stable
configurations, independently of the initial configuration. The moments of this measure,
in the thermodynamic limit $|\L| \to \infty$, are what we want to put in correspondence   
with the correlators of a conformal field theory. 

When the dynamics is started from a certain initial configuration, it produces at later
times two kinds of configurations, called transient and recurrent in the terminology of
Markov processes. The transient configurations are those which occur a finite number of
times only (they may not occur at all, depending on the initial configuration). In
the long run, they are not in the image of the dynamics, and have a zero measure with
respect to $P^{*}_{\L}$. A simple example is the configuration with all $h_{i}=1$, but
more generally, any configuration with two 1's at neighbour sites (sites $i,j$ with 
$\D_{ij}\neq 0$) is transient. 

The non transient configurations are recurrent. Their number is equal to $\det\D$, the
determinant of the toppling matrix, and asymptotically, they occur with equal probability,
so that the measure $P^{*}_{\L}$ is uniform on them \cite{dhar}. A criterion to decide
whether a given configuration is recurrent or transient is based on the notion of
forbidden sub--configuration (FSC): a sub--configuration, with support $\K \subset \L$,
is said to be forbidden if $h_{i}\le -\sum_{j \in \K\backslash\{i\}}\Delta_{ji}$,
for all sites $i$ of $\K$. For instance two neighbour sites with heights 1 form an FSC. Then
a configuration is recurrent if and only if it contains no FSC \cite{md2}. 

A practical way to test a configuration is to use the burning algorithm \cite{dhar}. At
time 0, all sites in $\L$ are unburnt and we define $\K_{0}=\L$ to be the set of unburnt
sites at time 0. The sites $i$ of $\K_{0}$ such that $h_{i} > -\sum_{j\in\K_0\backslash
\{i\}} \D_{ji}$ are burnable at time 1. So we burn them, obtaining a smaller set 
$\K_1 \subset \K_0$ of unburnt sites at time 1. Then the sites of $\K_{1}$ which are
burnable at time 2, {\it i.e.} those satisfying $h_{i} > -\sum_{j\in\K_1\backslash
\{i\}} \D_{ji}$, are burnt. This leaves a smaller set $\K_2 \subset \K_1$ of unburnt
sites at time 2. This burning process is carried on until no more site is burnable,
which means that $\K_{T+1}=\K_{T}$ for a certain $T$. Then the configuration is
recurrent if and only if all sites of $\L$ have been burnt ($\K_{T}=\emptyset$).
Otherwise $\K_{T}$ is an FSC.

The burning algorithm allows to define a unique rooted spanning tree on a graph
$\L^{\star}$, from the path followed by the fire in the lattice
\cite{md2}. The graph $\L^{\star}$ has the sites of $\L$ and the sink as vertices, and has
links defined by $\D$: an off--diagonal entry $\D_{ij} = -n$ means there are $n$ bonds
connecting the sites $i$ and $j$, and each site $i$ is connected to the sink by a number of
bonds equal to $\sum_{j \in \L} \D_{ij} \geq 0$, the number of grains dissipated when  
$i$ topples. At time 0, the sink is the only burnt site and forms the root of the tree. In
the next steps, the fire propagates from the sink to those sites which are burnable at time
1, then from the sites which have been burnt at time 1 to those which are burnable at time
2, and so on. If a site burns at time $t$, one says that it catches fire from one among its
neighbours that were burnt at time $t-1$ (or from the sink site at time 1). In case
there are more than one of these, a fixed ordering prescription is used to decide along
which bond the fire actually propagates. (The precise prescription will not be needed in
what follows; the interested reader is refered to \cite{md2} for an example of such a
prescription.) The collection of all bonds forming the fire path defines a spanning
tree, rooted in the sink, and growing towards the interior of the lattice $\L$.

This improved algorithm establishes a correspondence between the set of recurrent
configurations on $\L$ and the set of rooted spanning trees on $\L^{\star}$. The precise
mapping, although one--to--one, however depends on the prescription used. The prescription
we will use below differs slightly from the one defined in \cite{md2}, but is equally
valid.
The specific sandpile model we consider in the next sections is defined on the discrete 
upper half plane $\L = \Z_{>} \times \Z$, and has the massive discrete Laplacian as
toppling matrix subjected to the two different boundary conditions ``open'' and ``closed''
on the boundary, which we take to be the line $y=1$. The two toppling matrices are almost
identical and differ only along the boundary. They both depend on a positive parameter $t$,
controlling the rate at which sand is dissipated when sites topple. They read explicitely
\begin{equation}
\Delta^{\rm op}_{ij}=\cases{ 
4+t & if $i=j$,\cr
-1 &if $i$ and $j$ are n.n.,\cr
0 & otherwise,} \qquad
\Delta^{\rm cl}_{ij}=\cases{ 
4+t & if $i=j$ are off boundary,\cr
3+t & if $i=j$ are on boundary, \cr
-1 &if $i$ and $j$ are n.n.,\cr
0 & otherwise.}
\label{top}
\end{equation} 

One easily sees that $t$ grains of sand are dissipated (transfered to the sink) each time a
site topples, or $t+1$ if it is an open boundary site that topples. This model will be
called the massive Abelian sandpile model (MASM) in reference to the massive discrete
laplacian where $\sqrt t$ plays the role of a mass. The usual, critical model originally
defined in \cite{btw}, is recovered at $t=0$. 

In terms of the graph $\L^{\star}$, bulk sites and closed boundary sites have a $t$--fold
connection to the sink, while open boundary sites have a $(t+1)$--fold connection to it.
In addition, all nearest neighbour (n.n.) sites on $\L$ are connected by a single bond. 

An easy corollary of the above burning algorithm is that a site with height smaller or
equal to the number of its neighbours on $\L$ (3 or 4) is never burnt at time 1, and
therefore catches fire from one of its neighbours and not from the sink. Conversely, a site
with height strictly larger than $\D_{ii}(t=0)$ ---which we call a {\it supercritical
height}---, or an open boundary site with $h = 4$ is set afire by the sink. Supercritical
height values are those which exist only when $t$ is non zero. 

According to the definition of $\D$, $t$ should take integer values. However the MASM
correlations decay exponentially, with a correlation length that diverges only when $t$
goes to 0, like $1/\sqrt t$ \cite{tk,mr}. The large distance limit of the
lattice correlations must therefore be accompanied by a small $t$ limit, in such a way
that their scaling limit $\sqrt t = Ma \to 0,  |i-j| = |z|/a \to \infty$ be well
defined when $a \to 0$. So in practice, one expands the lattice MASM correlations in 
powers of $t$, and selects the dominant terms. These define correlators of a massive field
theory, which, in this case, turns out to be a massive perturbation \cite{mr} of $c=-2$ 
logarithmic conformal theory \cite{gur,k,gk},
\be
S={1\over\pi}\int (\p\t \bp \tt+M^2\t\tt/4),
\label{lag}
\ee
where $\t,\tt$ are anticommuting scalar fields. 

In the course of the calculations, we will make an extensive use of the inverse
toppling matrix $\D^{-1}_{ij}$. As is well--known, the inverse of $\D$ on the upper half
plane can be obtained in terms of the inverse massive Laplacian on the full plane $\D^{-1}$,
 via the image method. For $i=(m_{1},n_{1})$ and $j=(m_{2},n_{2})$, the explicit    
formulae read
\bea
&& \hspace{-7mm}(\D^{\rm op})^{-1}_{ij} = \D^{-1}_{ij} - \D^{-1}_{ij^{*}} 
= \D^{-1}_{ij} - \D^{-1}_{i^{*}j}\,, 
\qquad j^{*} = (m_{2},-n_{2}),\nonumber\\
&& \hspace{-7mm}(\D^{\rm cl})^{-1}_{ij} = \D^{-1}_{ij} + \D^{-1}_{i j^\vee} 
= \D^{-1}_{ij} + \D^{-1}_{i^\vee j}\,, 
\qquad  j^\vee = (m_{2},1-n_{2}).
\eea
The horizontal translation invariance is preserved in both cases, so that the entries of
the inverse matrices depend on $|m_{1}-m_{2}|, n_{1}$ and $n_{2}$. A short review on
values of the inverse massive Laplacian on the plane can be found in \cite{mr}.

The lattice open boundary condition is identified with the
Dirichlet condition in the continuum ($\t = \tt = 0$ on $\R$), whereas the closed  
boundary condition corresponds to the Neumann condition ($\partial \t -\bar      
\partial\t = \partial \tt - \bar \partial\tt = 0$ on $\R$). The Lagrangian 
(\ref{lag}) then implies the following Green functions on the upper half plane
\bea
&& \hspace{-7mm}\la \t(z) \t(w) \ra = \la \tt(z) \tt(w) \ra = 0,\\
&& \hspace{-7mm}\la \t(z) \tt(w) \ra_{\rm op} = 
K_{0}(M|z-w|) - K_{0}(M|z-\bar w|), \label{grop}\\
&& \hspace{-7mm}\la \t(z) \tt(w) \ra_{\rm cl} = 
K_{0}(M|z-w|) + K_{0}(M|z-\bar w|),
\eea
where $K_{0}$ is the modified Bessel function.


\section{Unit height and supercritical height variables}

Multisite probabilities for a number of sites to have height equal to 1 or supercritical
height values ($h_{i} > \D_{ii} - t$) is fairly easy if one uses the Bombay trick, a
beautiful technique designed by Majumdar and Dhar \cite{md1}. It can be formulated in
terms of height configurations or in terms of spanning trees. In this section, we will
use it in terms of heights, the formulation with trees being a particular case of the
general characterization given in the next section.

Suppose that we first want to compute the probability $P[h_{i_{0}}=1]$ that a certain site
$i_{0}$ has a height equal to 1. That probability is simply equal to the number of
recurrent configurations with a height 1 at $i_{0}$ divided by the total number of
recurrent configurations, which we know equals $\det \D$.

The idea of \cite{md1} is to define a new sandpile model in which the height at $i_{0}$
is always 1, and such that any recurrent configuration of this new model is in
correspondence with a recurrent configuration of the original model where the height at
$i_{0}$ is 1. To freeze the height at $i_{0}$ to the value 1, one simply reduces the
diagonal entry of the toppling matrix to 1. So the toppling matrix $\D'$ of the new
model will have $\D'_{i_{0}i_{0}} = 1$. Then in the new model, the site $i_{0}$ will
topple whenever its height exceeds 1, and each time it topples, it will lose a single grain
which will go to one the neighbours or to the sink. Consequently, $i_{0}$ will have a
single connection, either to the sink or to one its neighbours in $\L$. Finally, the
neighbours of $i_{0}$ cannot have a height equal to 1 in a recurrent configuration, so
that they assume only $\D_{ii}-1$ values. This can also be enforced in the new model by
decreasing the diagonal entries of $\D$ by 1 for those neighbours of $i_{0}$ which are
no longer connected to $i_{0}$. As the connections fix the off--diagonal part of the
toppling matrix, and the height ranges fix its diagonal part, this will determine $\D'$.

Thus the number of recurrent configurations with a height 1 at $i_{0}$ is equal to the
total number of recurrent configurations of the new model, itself equal to the determinant
of the new toppling matrix $\D'$. Setting $\D' = \D + B^{(i_{0})}$, one obtains  
\cite{md1}
\be
P[h_{i_{0}}=1] = {\det\D' \over \det\D} = \det(\I + \D^{-1}B^{(i_{0})}),
\ee
where $\D$ is the toppling matrix appropriate to the boundary condition one considers.
Because the difference $\D' - \D \equiv B^{(i_{0})}$ is non zero only on
sites around $i_{0}$, the previous formula reduces to the calculation of a finite
determinant, even on an unbounded lattice $\L$.

On the discrete upper half plane, the defect matrix $B^{(i_{0})}$ depends on the location
of $i_{0}$. If $i_{0}$ is off the boundary, and if one keeps it connected to one of its
four neighbours, then $B^{(i_{0})}$ is equal to
\be
B^{(i_{0})} = \pmatrix{-3-t & 1 & 1 & 1 \cr
1 & -1 & 0 & 0 \cr 1 & 0 & -1 & 0 \cr 1 & 0 & 0 & -1}
\ee
on $i_{0}$ (first label) and any three neighbours of $i_{0}$, and is identically zero
elsewhere (if the only connection of $i_{0}$ is to the sink, the $B$ matrix is    
5--by--5). In this case, the probability is given by a 4--by--4
determinant, and depends on the distance $m$ of $i_{0}$ to the boundary. At the critical
point ($t=0$) and for large values of $m$, it is equal to \cite{bip}
\be
P[h_{i_{0}}=1] = P_{1} \Big[1 \pm  {1 \over 4 m^2} + \ldots \Big],
\ee
where the $+$ (resp. $-$) sign refers to the open (resp. closed) boundary condition. 
$P_{1} = 2(\pi-2)/\pi^{3} = 0.0736$ is the probability that a site deep inside the
lattice (equivalently, on the infinite plane) has height 1 in the critical ASM.

If $i_{0}$ lies on the boundary, the matrix $B^{(i_{0})}$ depends on the boundary
condition and takes one of the two forms
\be
B^{(i_{0})}_{\rm op} = \pmatrix{-3-t & 1 & 1 \cr
1 & -1 & 0 \cr 1 & 0 & -1}, \qquad
B^{(i_{0})}_{\rm cl} = \pmatrix{-2-t & 1 & 1 \cr
1 & -1 & 0 \cr 1 & 0 & -1}.
\ee
The corresponding critical probability $P[h_{i_{0}}=1]$ is then a constant, which only
depends on the type of boundary the site $i_{0}$ is on \cite{iv},
\be
P_{1}^{\rm op} = {9 \over 2} - {42 \over \pi} + {320 \over 3\pi^{2}} - {512 \over
9\pi^{3}}, \qquad P_{1}^{\rm cl} = {3 \over 4} - {2 \over \pi}.
\ee
The one--site probability $P[h_{i_{0}}=1]$ can easily be computed for $t \neq 0$, but will
not be needed in what follows.

The probability that a site be supercritical can be treated in a similar way,
and is actually simpler. One now takes $t \neq 0$, since the probability does not make
sense at $t=0$.

Any site $i_{0}$, whatever its location and whatever the boundary condition, has $t$
possible supercritical height values, namely $h=\D_{i_0i_0}-t+1, \ldots, \D_{i_0i_0}$.
The probability that a site $i_{0}$ has a fixed supercritical height $h$ does
not depend on $h$, because a recurrent configuration remains recurrent if one replaces
a supercritical height at $i_{0}$ by another one. Therefore one has
\be
P[i_{0}\ {\rm is\ supercritical}] = P[h_{{i_{0}}} > \D_{i_{0}i_{0}} - t] = t \,
P[h_{i_{0}}=h],
\ee
where $h$ is any fixed supercritical value.

It is actually easier to compute the probability that $i_{0}$ is not supercritical. To
do that, one has to count the recurrent configurations with $h_{{i_{0}}} \leq
\D_{i_{0}i_{0}} - t$ ($=4$, or 3 on a closed boundary). In a new model defined by the 
new toppling matrix $\D'_{ij} = \D_{ij} - t \, \delta_{i,i_{0}}\,\delta_{j,i_{0}}$ on  
the same lattice, all recurrent configurations have $i_{0}$ not supercritical. One obtains 
\be
P[i_{0}\ {\rm is\ supercritical}] = 1 - {\det\D' \over \det\D} = t \,
\D^{-1}_{i_{0}i_{0}},
\label{super}
\ee
and 
\be
P[h_{i_{0}}=h] = \D^{-1}_{i_{0}i_{0}}.
\label{ssuper}
\ee
The defect matrix method works here too, and is simpler because the appropriate matrix
$\bar S^{(i_{0})}_{ij} = - t \, \delta_{i,i_{0}} \, \delta_{j,i_{0}}$ has rank 1. The
corresponding one--site probability is then given by a 1--by--1 determinant. 

Let us note that the probabilities (\ref{ssuper}) are well--defined for any strictly
positive value of $t$, but behave badly in the critical limit $t \to 0$. For a closed
boundary condition, they have a logarithmic singularity at $t=0$. For an open boundary
condition, they have a finite limit at $t=0$, but which is not a probability in general:
$\D^{-1}_{i_{0}i_{0}}=1-2/\pi=0.3634$  for $i_{0}$ on the boundary, and then grows
logarithmically with the distance of $i_{0}$ to the boundary.

So instead we will consider the probability (\ref{super}) for a site or a collection of
sites to be supercritical without specifying the actual heights. As we will see below,
that observable has well-defined correlations in the massive scaling limit, and
corresponds to a field that vanishes in the critical limit. 

After the one--site probabilities, multisite probabilities and correlations can be
computed by the same method almost routinely. The observables we consider in this
section are the two boundary random variables $\delta(h_{i} - 1)$ and
$\delta(i \ {\rm is\ supercritical})$ corresponding to the events ``$i$ has height 1'' or
``$i$ is supercritical'' for $i$ a site on the boundary of the upper half plane. In order
to get fields whose expectation value vanishes infinitely far from the boundary, one
considers the random variables subtracted by their average value. Anticipating the scaling
dimension 2 or 4 of the above random variables, we define their scaling fields by
\be
\phi^{\rm op,cl}_1(x) = \lim_{a \to 0} {1 \over a^{2}} \, [\delta(h_{x/a}-1) - 
P_{1}]\,, \quad
\phi^{\rm op,cl}_>(x) = \lim_{a \to 0} {1 \over a^{4,2}} \, [\delta(h_{x/a}\ 
{\rm is\ supercritical}) - \D^{-1}_{{x \over a}{x \over a}}],
\label{scale}
\ee
subjected to the scaling relations $t = a^2 M^2$ and $i\sqrt t = Mx$. 

To compute $n$--site probabilities, one simply inserts the proper defect matrices at the
locations of the observables, so that the full defect matrix is a direct sum of $n$
matrices $B^{(i)}$ or $\bar S^{(i)}$. One should however remember that $\bar S^{(i)}$ 
is not the defect matrix for $i$ being supercritical but for $i$ not being supercritical,
the complementary event. The scaling limit of the latter gives rise to a field
$\phi_{\leq}$, from which $\phi_{>} = -\phi_{\leq}$ is recovered. 
The probability then reduces to the calculation of a finite determinant whose entries are
combinations of entries of the inverse toppling matrix. As the scaling limit takes $t$ to
zero, one expands these entries in power series of $t$, keeping only the dominant term. 
>From (\ref{scale}) the latter yields the field theoretic correlation of fields $\phi_{1}$
and $\phi_{>}$, which are then identified with explicit fields of the Lagrangian theory
(\ref{lag}). This is a main advantage of working with the massive theory that this
identification is essentially unambiguous. 

The simplest way to proceed to the identification of the boundary fields $\phi_{1}$ and
$\phi_{>}$ is to use other lattice observables with already known field identifications.
Examples of such observables are precisely the bulk version of the above two random
variables. The corresponding bulk fields have been identified in \cite{mr} (see also
\cite{jeng2} for a proof that these identifications are consistent with a broad class of
multisite correlations),
\bea
\phi_1(z) &\!\!\!=\!\!\!& -P_{1} \, [\NP{\p\t \bp\tt + \bp\t \p\tt} + {M^{2} \over 
2\pi} \NP{\t\tt}],\\
\phi_>(z) &\!\!\!=\!\!\!& {M^2 \over 2\pi} \,\NP{\t\tt}.
\eea
One first computes the 2--point correlations involving one boundary observable and 
one bulk observable. From them, one may infer what the boundary fields must be, and then
cross--check their form from other correlations. 

We do not give much detail as the calculations are fairly straighforward, but simply
illustrate the method in a simple case, namely the identification of $\phi^{\rm
op}_{>}(x)$ on an open boundary.

We take two reference sites, $i_{0}$ on the boundary and $j_{0}$ in the bulk, far from
the boundary. The probability that they both be non supercritical reduces to a rank 2
determinant 
\be
P[i_{0}, j_{0}\ {\rm non\ supercrit.}] = {\det[\D^{\rm op} + \bar S^{(i_{0})} + 
\bar S^{(j_{0})}] \over \det \D^{\rm op}} =  \det\pmatrix{
1-t \, (\D^{\rm op})^{-1}_{i_{0}i_{0}} & -t \, (\D^{\rm op})^{-1}_{i_{0}j_{0}} \cr
-t \, (\D^{\rm op})^{-1}_{j_{0}i_{0}} & 1-t \, (\D^{\rm op})^{-1}_{j_{0}j_{0}}}.
\ee
Subtracting the product of disconnected probabilities obtained from (\ref{super}), one
has
\be
P[i_{0}, j_{0}\ {\rm non\ supercritical}]_{\rm connected} = -t^{2}\, 
(\D^{\rm op})^{-2}_{i_{0}j_{0}}.
\ee

If one chooses the two sites on a vertical line $i_{0} = (0,1)$ and $j_{0} = (0,m)$,
then (see for instance Appendix A of \cite{mr})
\be
(\D^{\rm op})^{-1}_{i_{0}j_{0}} = \D^{-1}_{(0,1)(0,m)} - \D^{-1}_{(0,-1)(0,m)} = 
-{\sqrt t \over \pi}\, K'_{0}(m\sqrt t) + \ldots
\ee
where the dots stand for subdominant terms in $t$. The dominant term in the connected 
2--site probability is thus
\be
P[i_{0}, j_{0}\ {\rm non\ supercritical}]_{\rm connected} = -{t^{3} \over \pi^{2}}\,
K_{0}'^{2}(m\sqrt t) + \ldots
\ee

Using the scaling relation (\ref{scale}), valid also for bulk fields (with a power
$a^{-2}$ for $\phi_{1}$ and $\phi_{\leq},\phi_{>}$), one finds the boundary/bulk 2--point
function
\be
\la \phi^{\rm op}_{\leq}(x) \, \phi_{\leq}(x+iy) \ra = -{M^{6} \over \pi^{2}}
K_{0}'^{2}(My).
\ee
>From the explicit form of $\phi_{\leq}(z) = -{M^2 \over 2\pi} \,\NP{\t\tt}$ given
above, one eventually arrives at
\be
\la \phi^{\rm op}_{\leq}(x) \,  \NP{\t\tt}(x+iy)\ra =  {2M^{4} \over \pi}
K_{0}'^{2}(My).
\ee

Using the Green function (\ref{grop}) on the upper half plane with an open boundary,
one sees that the only possible field assignment is $\phi^{\rm op}_{\leq}(x) = {2M^{2}
\over \pi} \NP{\p\t \p\tt}$, and therefore  $\phi^{\rm op}_{>}(x) = -{2M^{2} \over \pi}
\NP{\p\t \p\tt}$.

Proceeding in the same way for the other observables for the two boundary conditions, we
find the following scaling fields
\bea
\phi _{1}^{\rm op}  &\!\!=\!\!& \left({\frac{6}{\pi}-\frac{160}{3\pi^{2}}+
\frac{1024}{9\pi^{3}}}\right)  \NP{\p\t \p \tt}\,, \qquad 
\phi _{>}^{\rm op} =  -{2M^{2} \over \pi} \NP{\p\t \p \tt}\,,\\
\phi _{1}^{\rm cl} &\!\!=\!\!& -{\frac{8}{\pi}\Big(\frac{3}{4}-
\frac{2}{\pi }\Big)}\big[
\NP{\p\t\p\tt}+{\frac{1}{16}}M^{2} \NP{\t\tt} \big]\,, \qquad
\phi _{>}^{\rm cl} = {M^{2}\over 2\pi} \NP{\t\tt}\,.
\eea

These field identifications have been checked to be consistent with many multisite
probabilities: 2--site and 3--site boundary/boundary probabilities as well as 2--site
and 3--site mixed boundary/bulk probabilities. 
 
The massless limit is simply given by the limit $t\to 0$ in the MASM and $M\to 0$ in the
field theory. In this limit, the fields $\phi_{>}^{\rm op}$ and $\phi_{>}^{\rm cl}$ are
obviously null and the unit height fields for the two boundaries are identical up to a
numerical factor. One may note that the latters are proportional to the
holomorphic stress--energy tensor, and being descendants of the identity, they
belong to a chiral representation $\V_{0}$ \cite{gk}. This is consistent with the fact that
the only fields living on an open boundary are fields of $\V_{0}$, and that those living
on a closed boundary belong to an ${\mathcal R}_{0}$ representation, which contains
$\V_{0}$ as subrepresentation \cite{r,pr}.


\section{Spanning tree representation of recurrent configurations}

For the other height variables, the situation is not as easy. Although having a height 1 
or a height 2 at a given site does not seem to make much difference, the counting of the 
corresponding recurrent configurations is technically much more complicated for a height 
2 (or 3 or 4) than for a height 1. The defect matrix method no longer
works\footnote{Except a height 4 on an open boundary, which can be handled like a 
supercritical height.}, and the only
practical alternative seems to be the use of spanning trees. One then clearly sees the
difference: in terms of spanning trees, a height 1 is characterized by a local property of
the tree around the reference site, while the other heights are characterized by non local
properties of trees \cite{priez}.

As mentioned earlier, the rooted spanning tree is defined on $\L^{\star}$, the lattice
$\L$ augmented by the sink site, at which the tree is rooted. All sites $i$ of $\L$
are connected by $\Delta_{ii}$ bonds to other sites: $-\Delta_{ij}$ bond(s) connecting  
$i$ to $j$ and $\sum_{j\in\L}\Delta_{ij}$ bond(s) connecting $i$ to the root. With these
definitions, the Kirchhoff theorem asserts that the number of rooted spanning trees on
the graph $\L^{\star}$ defined by the matrix $\Delta$ is equal to $\N=\det\Delta$,
precisely the number of recurrent configurations. 

As a rooted spanning tree is a connected graph containing no loop, every site $i$ is
connected to the root by a unique path. A site $j$ is said to be a predecessor of the site
$i$ if the path from $j$ to the root passes through $i$, or equivalently, if $j$ lies on a
branch growing from $i$. A site $i$ which has no predecessor is called a leaf (the end of a
branch).

Priezzhev first and then Ivashkevich used the correspondence between recurrent
configurations and spanning trees to compute respectively the 1--site probabilities in the
plane, and the 1-- and 2--site probabilities on the boundary, open or closed, of the
upper half plane. For the 1--site probabilities at $i_{0}$, they decomposed the set of
recurrent configurations into subsets $S_{a}$, where $S_{a}$ contain the recurrent
configurations which remain allowed for any heights $h_{i_{0}}\geq a$ and which 
are forbidden otherwise. These subsets $S_{a}$ can be characterized in terms of rooted
spanning trees and their cardinal can be computed using classical results in graph theory.
As we will see, they decomposed local tree diagrams as sums of non local diagrams.
This system is invertible without further input for $i_{0}$ on a boundary, but is not
for $i_{0}$ in the bulk. So the calculation of probabilities for sites in the bulk is 
more complicated. 

For the 2--site probabilities, Ivashkevich used a similar decomposition of the recurrent
configurations into subsets $S_{ab}$. This decomposition however raises certain questions,
and will not be used here. Instead we set up a particular one--to--one map between
recurrent configurations and the rooted spanning trees, based on the burning algorithm. In
the case of single site probabilities, the mapping yields the same characterization in
terms of trees as the $S_{a}$ decomposition but is much more transparent in the case of
multisite probabilities. 

The burning algorithm, as we described it Section 2 complemented with an ordering
prescription, establishes a one--to--one mapping but with no clear correspondence
between the height values at the reference sites and the bond arrangements of the trees
around those sites (except for supercritical heights which are directly connected to the
root of the tree). For example, depending on the recurrent configuration, a site with a
height 4 can be a leaf on the tree or can support 1, 2 or 3 branches. To avoid this
problem, we proceed in two steps as follows, assuming that none of the reference sites is
supercritical. 

First, we run the burning algorithm and let the fire propagate through the lattice until no
more site is burnable but {\it preventing the reference sites from burning}. When this is
done, one is left with a sublattice $\L_{\rm b}$ of burnt sites and a complementary
sublattice $\L_{\rm u}$ of unburnt sites. The algorithm, using for example the ordering
prescription of \cite{md2}, yields at this stage the part of the spanning tree on $\L_{\rm
b}$. The other part $\L_{\rm u}$ is eventually burnable too and is actually burnt in the
second step. By definition, none of the reference sites is burnt yet, and a certain number
of them, at least one, are burnable. Those which are burnable are burnt simultaneously, and
trigger the fire propagation through $\L_{\rm u}$, thereby completing the spanning tree to
the whole lattice. So the complete tree is made of two pieces, a subtree $\T_{\rm b}$ on
$\L_{\rm b}$, and another $\T_{\rm u}$ on $\L_{\rm u}$. The subtree $\T_{\rm u}$ 
itself may have several roots which are among the reference sites those which were burnable
and which set fire to the whole of $\L_{\rm u}$. It is at those sites that $\T_{\rm u}$ is
grafted to $\T_{\rm b}$ to make the full tree $\T$. As we will see, only the shape of the
unburnt sublattice $\L_{\rm u}$ is used to characterize the height of the reference sites. 

This slightly modified burning algorithm establishes a well defined correspondence between
spanning trees and heights of the reference sites in the critical as well as in the massive
Abelian sandpile. Let us see how this works for the single--site probabilities, and how
it allows to compute the 2--site height correlations where one the two heights is equal
to 1 or is supercritical. At this stage we will be able to identify the boundary
fields corresponding to all heights. In the following section, we will compute other 2--site
and 3--site correlations to confirm these identifications.

Let us consider a configuration of the MASM on a square lattice $\L$, and let us focus on
 a fixed site $i_{0}$, the reference site. We will take $\L$ to be the upper half plane,
but what follows applies to any sort of portion of $\Z^{2}$, bounded or unbounded.
\begin{itemize}
\item 
If the height at $i_{0}$ is supercritical, then $i_{0}$ is set afire by the root (is
burnt at time 1). Thus all trees corresponding to those configurations have a bond
connecting the root and $i_{0}$. The probability that $i_{0}$ be supercritical is thus 
\begin{equation}
P[i_{0}\ {\rm is\ supercritical}] = {\N_{\star,i_{0}}\over\N},
\end{equation}
where $\N=\det\Delta$ and $\N_{\star,i_{0}}$ is the number of different spanning 
trees which ``use'' one of the $t$ bonds between the root and $i_{0}$. One way to
compute $\N_{\star,i_{0}}$ is to modify the toppling matrix by removing the connections
between $i_{0}$ and its nearest neighbors on $\L$ so that $i_{0}$ has connections only
to the root. Then $\N_{\star,i_{0}}=\det \D'$ with $\D'=\D+S^{(i_{0})}$ and the 
finite--dimensional defect matrix given by $S^{(i_{0})}_{i_{0}i_{0}}=t-
\D_{i_{0}i_{0}}$, $S^{(i_{0})}_{i_{0}i_{\ell}} = S^{(i_{0})}_{i_{\ell}i_{0}}=1$ 
for $i_{\ell}$ the nearest neighbors of $i_{0}$, and zero elsewhere. A simpler way is
however to compute $\N - \N_{\star,i_{0}}$, the number of trees which do not use the  
bonds between $i_{0}$ and the root. This can be done by removing precisely these bonds and
leads to the 1--dimensional defect matrix $\bar S^{(i_{0})}$ of the previous section,
with the result given in (\ref{super}).

The same arguments apply to a height 4 at an open boundary site. The only difference
is that such sites have $t+1$ connections to the root, so that
\begin{equation}
P_{4} = {1 \over t}\,{\N_{\star,i_{0}}\over\N} = \D^{-1}_{i_{0}i_{0}},
\qquad \mathrm{for\ }i_{0} \mathrm{\ on\ open\ boundary}.
\label{op4}
\end{equation}
\item 
If the height at $i_{0}$ is less or equal to the number of nearest neighbours on $\L$, we
use the burning algorithm to define a partition $\L = \L_{\rm b} \cup \L_{\rm u}$ as
explained above. As one looks here at 1--site probabilities, there is only one reference
site, so that in the sublattice $\L_{\rm u}=\L_{i_{0}}$, $i_{0}$ is the only burnable
site after all sites of $\L_{\rm b}$ have been burnt. It is therefore the root of the
subtree $\T_{\rm u} = \T_{i_{0}}$. The height at $i_{0}$ can now be related to the 
properties of the subtree $\T_{i_{0}}$. 

If $\L_{i_{0}}$ contains only the site $i_{0}$, then $h_{i_{0}}$ can take any of the 
values $1,2,\ldots,n_{i_{0}}$, where $n_{i_{0}}=3$ or 4 is the number of nearest 
neighbours of $i_{0}$. The full tree $\T$ is simply obtained by connecting $i_{0}$ to 
$\T_{\rm b}$ through one of the $n_{i_{0}}$ bonds connecting $i_{0}$ to its nearest
neighbours, so that $i_{0}$ is a leaf on $\T$ ($i_{0}$ must be connected to a nearest
neighbour in $\L$ and not to the root, since it catches fire from one of them).
If one denotes by $\N_{1}$ the number of spanning trees on $\L$ where $i_{0}$ is a leaf
grown on one its neighbours, then
\begin{equation} 
P_{1}={\N_{1}\over n_{i_{0}}\N}.
\label{p1}
\end{equation}

If $\L_{i_{0}}$ contains one nearest neighbor of the site $i_{0}$, the value of the
height at $i_{0}$ must be in the set $\{2,\ldots,n_{i_{0}}\}$ (it cannot be 1 since
otherwise $i_{0}$ would not be burnable and could not set $\L_{i_{0}}$ afire).
There are now $n_{i_{0}}-1$ possibilities to connect $\T_{i_{0}}$ to $T_{\rm b}$,
one for each nearest neighbour of $i_{0}$ in $\L_{\rm b}$. They correspond to the
height values compatible with the burning algorithm and thus the height 2 probability
reads
\begin{equation}
P_{2}={\N_{1}\over n_{i_{0}}\N}+{\N_{2}\over(n_{i_{0}}-1)\N},
\label{p2}
\end{equation}
where $\N_{2}$ is the number of spanning trees where $i_{0}$ has exactly one
predecessor among its nearest neighbours. 

The higher height probabilities can be determined by the same arguments,
\begin{equation}
P_{k}=P_{k-1}+{\N_{k}\over(n_{i_{0}}-k+1)\N},\quad P_{0}=0, 
\qquad k=1,\ldots,n_{i_{0}},
\label{pk}
\end{equation}
where $\N_{k}$ is the number of spanning trees $\T$ on $\L$ in which $i_{0}$ has exactly
$k-1$ predecessors among his neighbours.
\end{itemize}

One sees that the computation of the various 1--site probabilities requires the calculation
of the numbers $\N_{k}$. If the lattice $\L$ is the discrete upper half plane, and for
$i_{0}$ a site on the boundary, the Figure 1 describes the types of trees which
contribute to the different $\N_{k}$'s.

\begin{figure}[htb]
\psfrag{N1}{$\N_1\,=$}
\psfrag{N2}{$\N_2\,=$}
\psfrag{N3}{$\N_3\,=$}
\psfrag{p}{$+$}
\psfrag{a1}{$\alpha_1$}
\psfrag{a2}{$\alpha_2$}
\psfrag{a3}{$\alpha_3$}
\psfrag{b1}{$\beta_1$}
\psfrag{b2}{$\beta_2$}
\psfrag{tb1}{$\tilde\beta_1$}
\psfrag{tb2}{$\tilde\beta_2$}
\psfrag{g1}{$\gamma_1$}
\psfrag{g2}{$\gamma_2$}
\psfrag{d}{$\delta$}
\psfrag{e}{$\varepsilon$}
\psfrag{te}{$\tilde\varepsilon$}
\psfrag{fa}{$\phi_a$}
\psfrag{fb}{$\phi_b$}
\psfrag{tfa}{$\tilde\phi_a$}
\psfrag{tfb}{$\tilde\phi_b$}
\center{\mbox{\includegraphics[scale=.33]{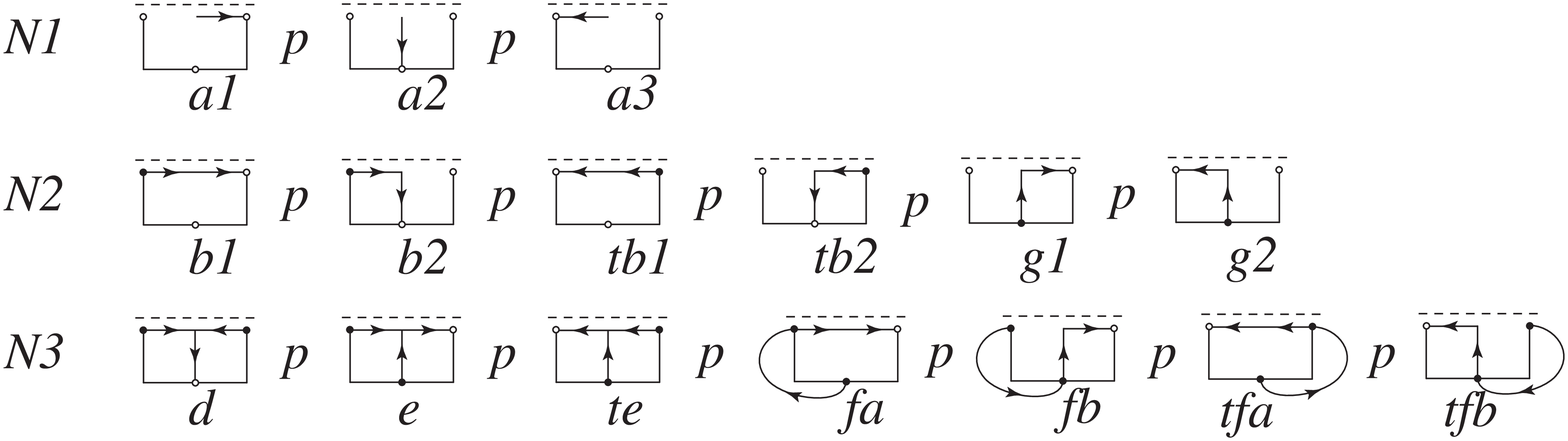}}}
\caption{\small Non local diagrams contributing to the 1--site probabilities.}
\end{figure}

These diagrams represent the restriction of trees to four sites, namely the reference 
site $i_{0}$ and its three neighbours (the dashed line represents the border of $\L$,
pictured as the lower half plane !). The arrows indicate the direction of the path towards
the root (opposite to the fire propagation line). The black dots are the nearest neighbours
which are predecessors of $i_{0}$, the
white dots are those which are not. Those diagrams labelled by identical greek letters
contribute the same amount to the corresponding $\N_{k}$. The tilded letters refer to
diagrams which are the mirror images of the diagrams with untilded letters.

The diagrams in Figure 1 represent non local constraints on the compatible trees.
The presence of an arrow between $i_{0}$ and one or more neighbours poses no  
computational problem, because it only means that the tree has to use specific bonds,
and the counting of those trees can be achieved by modifying {\it locally} the toppling
matrix by a finite rank defect matrix. But for a neighbour to be a predecessor or not is a
non local property, and enumerating the relevant trees is trickier. For the trees
contributing to $\N_{3}$ for instance, one sees that a nearest neighbour, call it
$i_{1}$, of $i_{0}$ can be a predecessor of $i_{0}$ because
$i_{1}$ is connected to $i_{0}$ through the nearest neighbour bond ($i_{1}$ catches fire
directly from $i_{0}$, like in $\delta, \epsilon$ and $\tilde \epsilon$), or through a 
long path around the lattice ($i_{1}$ catches fire after a long sequence of burnings, long
after $i_{0}$ got burnt, like in $\phi$ and $\tilde \phi$).

Various classes of non local diagrams in Figure 1, with their constraints on
predecessorships, sum up to local diagrams, where these constraint are relaxed and only
local arrow configurations are imposed. As the local diagrams are easily calculable by
toppling matrix adjustments, this yields linear relations among the non local
contributions. For a generic position of the reference site $i_{0}$, the linear system is
not invertible and is not sufficient to compute the non local contributions. The crucial
observation made by Ivashkevich in \cite{iv} was to note that it actually becomes 
invertible if $i_{0}$ is on a boundary, which allows to reduce the non local
contributions to local calculations (this statement will have to be qualified in the
case of 2--site insertions). Indeed the relations read explicitely \cite{iv} (as we
consider the heights 1, 2 and 3 only, we do not need to distinguish the diagrams for closed
and open boundaries, which only make a difference through the proper toppling matrix to
be used in the explicit local computations; at the same time, that also makes the number of
equations smaller) 
{\newcommand\fig[1]{
\psfrag{P}{$Q_0\,=$}
\psfrag{Pa}{$Q_1\,=$}
\psfrag{Pat}{$\tilde Q_1\,=$}
\psfrag{Pb}{$Q_2\,=$}
\psfrag{Pbt}{$\tilde Q_2\,=$}
\psfrag{Pc}{$Q_3\,=$}
\psfrag{Pct}{$\tilde Q_3\,=$}
\psfrag{Pd}{$Q_4\,=$}
\psfrag{Pe}{$Q_5\,=$}
\psfrag{Pet}{$\tilde Q_5\,=$}
\psfrag{p}{$+$}
\psfrag{e}{$=$}
\psfrag{a}{$\alpha$}
\psfrag{b}{$\beta$}
\psfrag{tb}{$\tilde\beta$}
\psfrag{g}{$\gamma$}
\psfrag{d}{$\delta$}
\psfrag{eps}{$\varepsilon$}
\psfrag{teps}{$\tilde\varepsilon$}
\psfrag{f}{$\phi$}
\psfrag{tf}{$\tilde\phi$}
\parbox{70mm}{\includegraphics[scale=.33]{diag/#1}}}
\begin{equation}
\fig{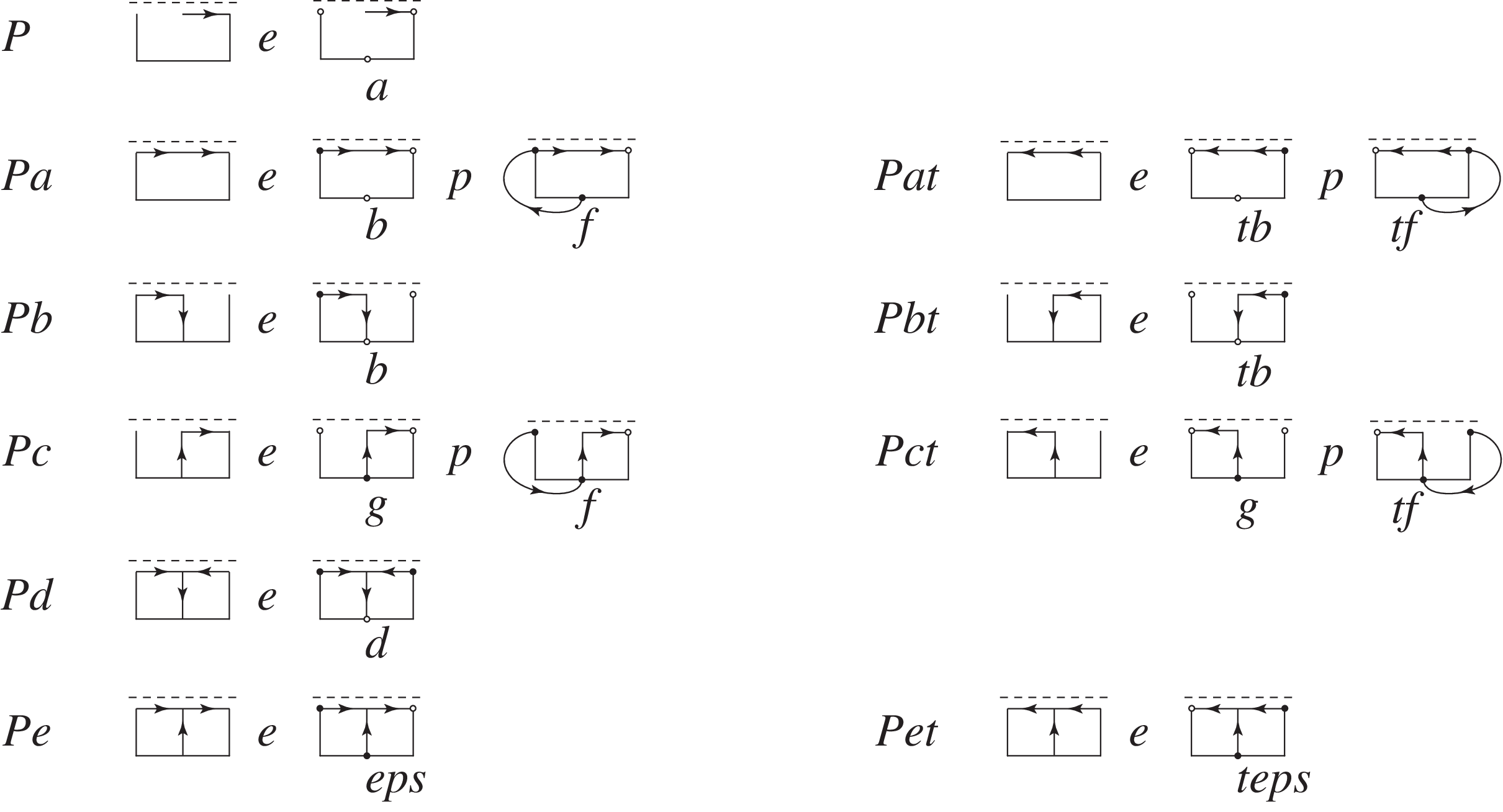}
\label{pi}
\end{equation}}
The four equations on the right are clearly obtained from the corresponding four on the
left by a mirror symmetry about the reference site $i_{0}$. As the toppling matrix is
invariant under that symmetry, the three tilded diagrams contribute the same amount as
the untilded ones, $\tilde\beta = \beta, \tilde \varepsilon=\varepsilon, 
\tilde\phi=\phi$, as do the tilded and the untilded local diagrams. Thus the four
equations on the right are redundant, and one is left with the linear system on the left.
This system is manifestly invertible for the six non local contributions, noted $\alpha,
\beta, \gamma, \delta, \varepsilon$ and $\phi$.

Let us note that the non local diagrams $\alpha, \beta, \delta$ and $\varepsilon$ turn  
out to be entirely local, because the arrow configurations make the predecessorship
properties redundant. As the height 1 probability $P_{1}$ is given solely in terms of
$\alpha$, its computation is purely local. This remains true for any multisite height 1
probabilities and for arbitrary positions, in the bulk or on boundaries. 

The Kirchoff theorem allows the local diagrams to be computed by the defect matrix  
method. The presence resp. the absence of an arrow from $i$ to $j$ means that one counts
all trees which contain resp. do not contain that oriented bond. One defines a
new toppling matrix $\D'$ by setting to $-\epsilon$ the $i,j$ entry if the $i \to j$ bond
is to be used in the tree, and to 0 if that bond is not to be used; moreover the diagonal
entries of $\D'_{ii} = \D_{ii}$ should remain equal to the number of bonds
going out from $i$. Then if $n$ bonds are to be used, the determinant of $\D'$ will contain
a highest degree term $\epsilon^{n}$ whose coefficient is the number of trees which
precisely use the given $n$ bonds in the prescribed direction (see for instance
\cite{priez}).

Writing as before $\D' = \D + B$, one finds for $Q_4$ for example 
\be
B = \pmatrix{-3 +\epsilon &  1 &  1 - \epsilon &  1 \cr 
1 - \epsilon & -1 + \epsilon &  0 & 0 \cr 
1 &  0 &  -1 &  0 \cr 
1 - \epsilon & 0 &  0 &  -1 +\epsilon},
\label{Beps}
\ee
and, in the critical limit,
\be
{Q_4 \over \N} = {\delta \over \N} = \lim_{\epsilon \to \infty} {1 \over \epsilon^{3}} \, 
\det[\I + \D^{-1}B] = \cases{{1 \over \pi} - {1 \over 4} & on closed boundary, \cr
\noalign{\smallskip}
{(3\pi - 8)^{3} \over 9\pi^{3}} & on open boundary.}
\ee
The calculation of the other five local diagrams and then the inversion of the linear
system yields the values of $\N_{1}, \N_{2}$ and $\N_{3}$, and in turn of $P_{1},
P_{2}$ and $P_{3}$. In the critical limit, one recovers the numbers given in \cite{iv}.

In order to identify the height boundary fields, we need 2--site correlations involving
the boundary heights 2 and 3. Again the simplest is to look at the correlations of a
boundary height 2 or 3 with a known boundary variable, namely a height 1 or a 
supercritical height value. The advantage is that the latters are already known from the
previous section, but more importantly, they correspond to local defect matrix insertions.
This makes the above formalism, useful to compute 1--site probabilities, essentially valid.

Because one can force a site $i_{0}$ to have height 1 or to be supercritical by modifying
the toppling matrix by $\D \longrightarrow \D(i_{0}) = \D + B^{(i_{0})}$ or $\D +
S^{(i_{0})}$ (or better $\D + \bar S^{(i_{0})}$), the 2--site probabilities 
$P[h_{i_{0}}=1\ {\rm or\ supercr.}, h_{j_{0}}=
2\ {\rm or\ }3]$ can be viewed as 1--site probabilities for the height at $j_{0}$ but
with the toppling matrix $\D(i_{0})$ to account for the constraint at $i_{0}$. Then the
above method remains completely valid provided we replace $\D$ by the appropriate
$\D(i_{0})$, itself to be modified by matrices like in (\ref{Beps}) in order to compute
local diagrams \footnote{To keep the decompositions of the $\N_{k}$ in terms of the non
local diagrams as in Figure 1, the site $i_{0}$ should not be too close to $j_{0}$.}.
If one does that and use $\D(i_{0})$ as the normalizing toppling matrix,
one is actually computing the conditional probability for having a 2 or a 3 at $j_{0}$
conditioned on having a height 1 or a supercritical height at $i_{0}$. To get the joint
probabilities, one simply multiplies the final answers by $P[h_{i_{0}}=1]$ or 
$P[h_{i_{0}}={\rm supercr.}]$.

The non local diagrams contributing to the numbers $\N_{k}$ remain as in Figure 1.
However the tilded and the untilded diagrams no longer contribute equally because the
mirror image about $j_{0}$ spoils the constraint at $i_{0}$, and does not leave the
toppling matrix $\D(i_{0})$ invariant. Therefore it is the full system (\ref{pi}) that
needs be solved. It is overdetermined as it involves 10 equations for only 9
unknowns, but the number of equations is reduced by one due to the following identity
\be
\tilde P(1) + P(2) + P(3) = P(1) + \tilde P(2) + \tilde P(3),
\label{constr}
\ee
satisfied for all values of $t$ as a simple consequence of the fact that the inverse of
$\D(i_{0})$ satisfies a discrete Poisson equation. The procedure is otherwise identical to
that for the 1--site probabilities.
     
For the open boundary, the boundary joint probabilities of a site with height 2, 3 or
4 and a site with height 1 or with a supercritical height all have the same form as 
two unit heights on the boundary. It means the same field identification up to a
numerical factor:
\bea
\phi_{2}^{\rm op}&=&\textstyle\big(-{18\over\pi}+{400\over3\pi^2}-
{2048\over9\pi^3}\big)\NP{\p\t\p\tt},\\
\phi_{3}^{\rm op}&=&\textstyle\big({14\over\pi}-{80\over\pi^2}+
{1024\over9\pi^3}\big) \NP{\p\t\p\tt},\\
\phi_{4}^{\rm op}&=&\textstyle-{2\over\pi} \NP{\p\t\p\tt}.
\eea
The last line can most easily obtained from (\ref{op4}), which implies 
$\phi^{\rm op}_{4} = {1 \over M^{2}} \phi^{\rm op}_{>}$. 

For the closed boundary, the correlations involving a height 2 or 3 have a more 
complicated structure. For example, one finds (with $m=|i_{0}-j_{0}|$)
\bea
&& \hspace{-8mm} P[h_{i_{0}}>3, \, h_{j_{0}}=2]=\textstyle\frac{t^{2}}
{\pi^{2}}\Big[\frac{2}{\pi}K^{2}_{0}(m\sqrt{t})-(3-\frac{12}
{\pi})K'^{2}_{0}(m\sqrt{t})-\frac{1}{2}K_{0}(m\sqrt{t})
K_{0}''(m\sqrt{t})\Big]+\ldots \\
&& \hspace{-8mm} P[h_{i_{0}}>3, \, h_{j_{0}}=3]=\textstyle\frac{t^{2}}
{\pi^{2}}\Big[\frac{1}{4}K^{2}_{0}(m\sqrt{t})-\frac{4}{\pi }
K'^{2}_{0}(m\sqrt{t})+\frac{1}{2}K_{0}(m\sqrt{t})
K_{0}''(m\sqrt{t})\Big]+\ldots
\eea
These results and the corresponding ones for $h_{i_{0}}=1$ are compatible with the
following field assignments for the height 2 and 3
\bea
\phi_{2}^{\rm cl} &\!\!=\!\!& \textstyle\big({6\over\pi}-{24\over\pi^2}\big)
\NP{\p\t\p\tt}+{1\over2\pi}\NP{\t\p\p\tt}
+\big({1\over8\pi}-{1\over\pi^2}\big) M^2 \NP{\t\tt},\\
\phi_{3}^{\rm  cl} &\!\!=\!\!& \textstyle{8\over\pi^2}\NP{\p\t\p\tt}-
{1\over2\pi}\NP{\t\p\p\tt}-{1\over4\pi} M^2 \NP{\t\tt}.
\eea
This identification is in this case not unique, since the field theory is invariant under
$\t \to \tt$ and $\tt \to -\t$. One could in particular change $\NP{\t(\p\p\tt)}$ for 
$\NP{(\p\p\t)\tt}$, which are different fields since their correlation contains
a logarithm while their self--correlations do not. If one requires that the sum 
$\phi^{\rm cl}_{1} + \phi^{\rm cl}_{2} + \phi^{\rm cl}_{3}$ be zero in the massless 
limit, then the choice for one the two fields must be made for both $\phi^{\rm cl}_{2}$ 
and $\phi^{\rm cl}_{3}$. The correlations computed in the next section confirm this.
Note that the sum of fields $\phi^{\rm cl}_{1} + \phi^{\rm cl}_{2} + \phi^{\rm cl}_{3} + 
\phi^{\rm cl}_{>}$ vanishes identically. The similar sum in the open case, $\phi^{\rm
op}_{1} + \phi^{\rm op}_{2} + \phi^{\rm op}_{3} + \phi^{\rm op}_{4} + 
\phi^{\rm op}_{>}$, vanishes at the critical point only, because the dimension of
$\phi^{\rm op}_{>}$ does not match the dimension of the universal terms of the other
fields.

At this stage, all boundary height fields for the massive Abelian sandpile model have
been determined. To have more checks on the field identifications, we compute
in the following section, all 2--site and some 3--site height correlations.


\section{Higher boundary correlations}

In the previous section we have seen that the multisite probabilities where only one
reference site $i_{0}$ has a height value in $\{2,3\}$ can be computed from the
diagrams listed in Figure 1 by using a toppling matrix properly decorated by defect
matrices to account for height constraints (height 1 or supercritical) at the other
sites. The calculation of multisite probabilities where two reference sites $i_{0}$ and
$j_{0}$ have a height value in $\{2,3\}$ leads naturally to pairs of such 
diagrams, one at $i_{0}$, the other at $j_{0}$. However the situation becomes technically 
more complex because sites in the diagram at $i_{0}$ can be predecessors of $j_{0}$
and/or the other way round. So the topology of the spanning trees can be more
complicated and their counting more difficult. 

Let us first consider the 2--site probabilities $P_{ab} = P[h_{i_{0}}=a,\, h_{j_{0}}=b]$
for $a,b$ in $\{2,3\}$ and where $i_{0}$ and $j_{0}$ are on the boundary of the upper 
half plane. We start the burning algorithm as explained in Section 4 without ever
burning the sites $i_{0}$ and $j_{0}$, and until no other sites than those two are
burnable. This yields a sublattice $\L_{\rm u}=\L_{i_{0}j_{0}}$ of unburnt sites, which 
subsequently catches fire either from $i_{0}$ or from $j_{0}$, or from both if they are
both burnable. In turn the fire propagation on $\L_{i_{0}j_{0}}$ defines a subtree
$\T_{\rm u} = \T_{i_{0}j_{0}}$, rooted at $i_{0}$, or at $j_{0}$, or at both sites. The
full tree $\T$ is made up of the subtree $\T_{\rm b}$ living on the sublattice of burnt
sites, to which $\T_{i_{0}j_{0}}$ is grafted at $i_{0}$ and/or $j_{0}$.

The restriction of any tree to the neighbourhood of a reference site looks like one of
the non local diagrams shown in Figure 1. So one can visualize the restriction to the two
neighbourhoods by a pair of such diagrams. Using the same labelling as in
Figure 1, we will denote the pairs of diagrams by pairs of greek letters (with indices),
the first one for the diagram around $i_{0}$, the other for the diagram at $j_{0}$. In an
obvious notation, a pair of greek letters belongs to a certain set $\N_{k} \times \N_{l}$.
As we did in Section 4 for the one--site probabilities, we have to compute which
probabilities $P_{ab}$ a pair of diagrams contributes to. 

For 1--site probabilities, we know from Section 4 that the diagrams in $\N_{k}$
contribute equally to the probabilities $P_{a}$ for $k \leq a \leq 3$. Indeed the
three diagrams $\alpha_{1},\, \alpha_{2},\, \alpha_{3}$ of $\N_{1}$ are obtained 
from each other by changing the arrow around the reference site. The change 
converts a tree which is compatible with a diagram $\alpha_{i}$ into a tree which is
compatible with another diagram $\alpha_{j}$, and this fact shows that the number of 
trees compatible with a diagram $\alpha_{i}$ does not depend on $i$, namely $\alpha_1  
= \alpha_2 = \alpha_3$ or $\N_1 = 3\alpha$. As the position of the arrow determines
univoquely the height value, the three probabilities $P_{1},\, P_{2},\, P_{3}$ get an 
equal contribution $\N_{1}/3\N$ from the diagrams in $\N_{1}$. The same is true of the 
six diagrams in  $\N_{2}$. They come in pairs ($\beta_{1},\beta_{2}$), ($\tilde
\beta_{1},\tilde  \beta_{2}$), ($\gamma_{1},\gamma_{2}$), where the diagrams 
within a pair are related by changing the direction of the arrow coming out from the
reference site. The same arguments as above show that $\beta_{1} = \beta_{2}$, $\tilde
\beta_{1} = \tilde  \beta_{2}$, $\gamma_{1} = \gamma_{2}$,  and that $P_{2}$, 
$P_{3}$ receive an identical contribution $\beta + \tilde\beta + \gamma = \N_{2}/2\N$ 
from the diagrams in $\N_{2}$. For $\N_{3}$, each diagram is on its own and contributes 
to $P_{3}$. 

In the case  of 2--site probabilities, the same arguments would show that the diagrams in
$\N_{k} \times \N_{l}$ contribute equally to the probabilities $P_{ab}$ for 
$k \leq a \leq 3$, $l \leq b \leq 3$, provided one can prove that changing the direction
of an arrow in the way recalled above in either diagram, or in both diagrams, turns a
compatible tree into a compatible tree of the same class. Because the two diagrams can now
be linked by fire paths, this is no longer guaranteed, and actually fails in a few
cases, pictured in Figure 2.

On the first line of Figure 2, one sees for instance that the diagram denoted by $A_{3}$
is a pair $\gamma_2\beta_1$. It is linked in such a way that when one changes the arrow in
$\beta_{1}$, one obtains a well--defined tree (noted $A_{2}$) 
compatible with a pair $\tilde\phi_{b}\beta_{2}$. If one changes in 
$A_{3}$ the arrow of $\gamma_{2}$, one obtains the diagram $A_{1}$, of the  
type $\gamma_{1}\phi_{a}$. Changing the arrow of $\gamma_{2}$ and of $\beta_{1}$
introduces a loop, and so cannot contribute to a 2--site probability. The trees
compatible with the diagrams $A_{1}$, $A_{2}$ and $A_{3}$ are related by local changes
of arrow, but belong to different classes, namely $\N_{2} \times \N_{3}$,  $\N_{3} 
\times \N_{2}$ and $\N_{2} \times \N_{2}$. There should normally be a fourth 
diagram, in $\N_{3} \times \N_{3}$, but which does not exist as a tree. 

It is not difficult to see that the misbehaviours with respect to arrow changes can only
be of the type shown by the triplet $(A_{1},A_{2},A_{3})$. When the two 
diagrams are tied in a special way by the fire paths, one change of arrow in a
diagram in $\N_{2} \times \N_{2}$ sends it to a diagram in $\N_2 \times \N_3$ or
$\N_{3} \times \N_{2}$, and two arrow changes introduce a loop. 

Figure 2 shows four triplets of diagrams where this peculiar behaviour occurs.
Diagrams labelled by the same capital letter are in equal number, since the numbers of
compatible trees are equal. The twelve diagrams shown in Figure 2 and the mirror 
diagrams (not shown in Figure 2), obtained by exchanging the diagram at $i_{0}$ with the
reflected one at $j_{0}$ and vice--versa, make the complete list of misbehaved diagrams.
We will denote the mirror diagrams with tildes.

\begin{figure}[htb]
\psfrag{A1}{$A_1\,=$}
\psfrag{A2}{$A_2\,=$}
\psfrag{A3}{$A_3\,=$}
\psfrag{B1}{$B_1\,=$}
\psfrag{B2}{$B_2\,=$}
\psfrag{B3}{$B_3\,=$}
\psfrag{C1}{$C_1\,=$}
\psfrag{C2}{$C_2\,=$}
\psfrag{C3}{$C_3\,=$}
\psfrag{C4}{$C_4\,=$}
\psfrag{C5}{$C_5\,=$}
\psfrag{C6}{$C_6\,=$}
\center{\mbox{\includegraphics[scale=.33]{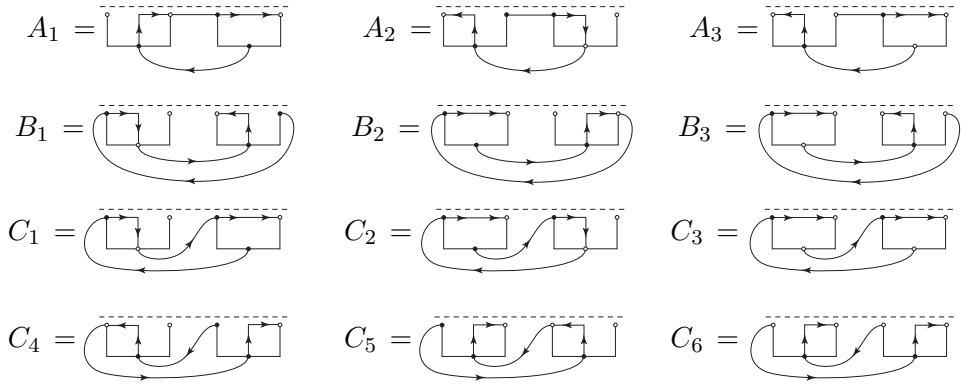}}}
\caption{\small Non local diagrams representing spanning trees which have an 
anomalous behaviour under a local change of arrow around $i_{0}$ and/or $j_{0}$. The 
mirror diagrams must be added to have the complete list of such diagrams.}
\end{figure}

The two--step burning algorithm allows to determine which probability each diagram
contributes to. In the diagram $A_{1}$ for instance, the subtree $\T_{i_{0}j_{0}}$
catches fire from the eastern neighbour of $j_{0}$. It is thus burnable at a time where
only one of its neighbours is burnt, and so must have a height 3. The other reference
site $i_{0}$ is not burnable at the time $\T_{i_{0}j_{0}}$ catches fire despite the
fact that its western neighbour was burnt, which implies that its height is at most 2. When
$i_{0}$ is burnable, it has two burnt neighbours and one southern
unburnt neighbour, meaning that its height must be 2. Thus $A_{1}$ contributes to 
$P_{23}$. One finds similarly that the first column in Figure 2 contribute to $P_{23}$, the
second column to $P_{32}$, and the last column to $P_{33}$.

We define $[\N_{k} \times \N_{l}]$ to be the set of trees in $\N_{k} \times
\N_{l}$ which do not have this sort of misbehaviour under a local change of
arrow. The set $\N_{k} \times \N_{l}$ is equal to $[\N_{k} \times \N_{l}]$
except in the following three cases,
\bea
\N_{2} \times \N_{2} \egal [\N_{2} \times \N_{2}] + A_{3} + B_{3} + C_{3} + 
C_{6} + \tilde A_{3} + \tilde B_{3} + \tilde C_{3} + \tilde C_{6},\\
\N_{2} \times \N_{3} \egal [\N_{2} \times \N_{3}] + A_{1} + B_{1} + 
C_{1} + C_{4} + \tilde A_{2} +  \tilde B_{2} + \tilde C_{2} + \tilde C_{5},\\
\N_{3} \times \N_{2} \egal [\N_{3} \times \N_{2}] + A_{2} + B_{2} + 
C_{2} + C_{5} + \tilde A_{1} + \tilde B_{1} + \tilde C_{1} + \tilde C_{4}.
\eea

The trees in $[\N_{k} \times \N_{l}]$ contribute equally to the probabilities
$P_{ab}$, $k \leq a \leq 3$ and $l \leq b \leq 3$, while those compatible with
the diagrams of Figure 2 must be handled separately. One obtains
\bea
P_{22} \egal P_{12} + P_{21} - P_{11} + {[\N_{2} \times \N_{2}] \over 4\N} ,\\
P_{23} \egal P_{13} + P_{22} - P_{12} + {[\N_{2} \times \N_{3}] \over 2\N} + 
{1 \over \N} [A_{1} + B_{1} + C_{1} + C_{4} + \tilde A_{2} +  \tilde B_{2} + 
\tilde C_{2} + \tilde C_{5}],\\
P_{32} \egal P_{22} + P_{31} - P_{21} + {[\N_{3} \times \N_{2}] \over 2\N} + 
{1 \over \N} [A_{2} + B_{2} +  C_{2} + C_{5} + \tilde A_{1} + \tilde B_{1} + 
\tilde C_{1} + \tilde C_{4}],\\
P_{33} \egal P_{23} + P_{32} - P_{22} + {\N_{3} \times \N_{3} \over \N} + 
{1 \over \N} [A_{3} + B_{3} + C_{3} + C_{6} + \tilde A_{3} + \tilde B_{3} + 
\tilde C_{3} + \tilde C_{6}] \nonumber\\
&& \hspace{1cm} - {1 \over \N} [A_{1} + A_{2} + B_{1} + B_{2} + 
C_{1} + C_{2} + C_{4} + C_{5} + {\rm mirrors}].
\eea
The subtracted term in $P_{33}$ is due to the fact that the part of $P_{23},P_{32}$
related to the misbehaved diagrams in Figure 2 (first and second columns) do not
contribute to $P_{33}$.

The sets $[\N_{k} \times \N_{l}]$ will be further partitioned in classes
labelled by a pair of diagrams, {\it f.i.} $[\N_{2} \times \N_{2}] =
[\beta_1\beta_1] + [\beta_{1}\beta_{2}] + \ldots$. One will remember 
that the cardinal of a class does not depend on the {\it numerical} indices
attached to diagrams, so that $|[\beta_{1}\beta_{1}]| = |[\beta_{1}\beta_{2}]| =
|[\beta_{2}\beta_{1}]| = |[\beta_{2}\beta_{2}]|$, and so on. Replacing however
$\phi_{a}$ by $\phi_{b}$ in a class does not necessarily conserve the
cardinal of that class, {\it f.i.} $|[\beta\phi_{a}]| \neq |[\beta\phi_{b}]|$.
The 2--site probabilities can be computed if the numbers of trees in these subclasses and
of those compatible with the non local diagrams of Figure 2 can be calculated. 

As for the 1--site probabilities, we can decompose each pair of local diagrams as a sum
of non local ones. We have for example, 
\newcommand\figA{\parbox{115mm}{
\psfrag{Paa}{$Q_{1,1}\,=$}
\psfrag{e}{$=$}
\psfrag{p}{$+$}
\psfrag{betbet}{$[\beta\beta]$}
\psfrag{phiphi}{$\phi_a\phi_a$}
\psfrag{betaphi}{$[\beta\phi_a]$}
\psfrag{phibeta}{$[\phi_a\beta]$}
\psfrag{C}{$C_3$}
\includegraphics[scale=.33]{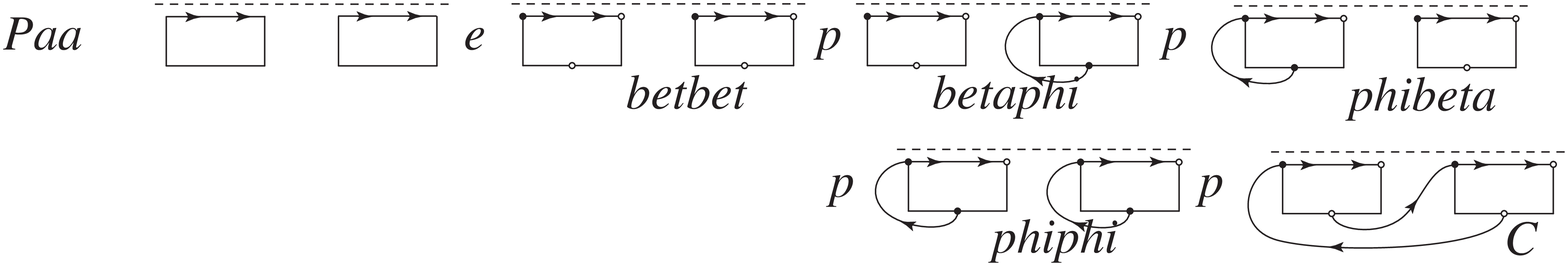}}}
\newcommand\figB{\parbox{115mm}{
\psfrag{Pbc}{$Q_{2,3}\,=$}
\psfrag{e}{$=$}
\psfrag{p}{$+$}
\psfrag{betagam}{$[\beta\gamma]$}
\psfrag{betaphi}{$[\beta\phi_b]$}
\includegraphics[scale=.33]{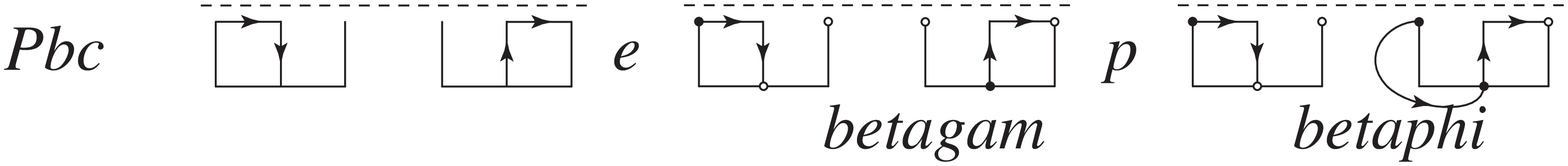}}}
\bea
&&\figA \label{q11}\\
&&\figB \label{q23}
\eea

As such, the linear system one obtains in this way is underdetermined.
Let us proceed to the counting in the general case, that is, when heights equal to 1 or 
supercritical heights are inserted at other places than $i_{0}$ and $j_{0}$. In this
situation, the full system must include all pairs of local diagrams (for example, one
would have the equation $Q_{1,1}$ and its mirror image $\tilde Q_{1,1}$). 

There are 81 equations like (\ref{q11}) and (\ref{q23}), since every such equation
is a pair of local diagrams, chosen from the nine diagrams appearing on the
last five lines of Eq. (\ref{pi}). There are 9 independent variables for the classes of
$[\N_{2} \times \N_{2}]$ (pairs of
elements in $\{\beta, \tilde\beta,\gamma\}$ since the indices are irrelevant), 
21 variables for the classes of $[\N_{2} \times \N_{3}]$ and $[\N_3 \times \N_2]$,
and 49 variables for $\N_{3} \times \N_{3}$. To these one must add the four   
variables $A$, $B$, $C$ and $\tilde C$, for the diagrams of Figure 2 (one can show that
$A = \tilde A$ and $B = \tilde B$).

In total, one has a linear system of 81 equations for 104 variables. It is actually worse 
because the equations are not all independent, due to some non trivial identities
among local diagrams (like in (\ref{constr})). It is however possible to compute 
the probabilities in terms of a reduced number of variables.

In the calculation of the 1--site probabilities, the non local diagrams $\phi_{a}$ and
$\phi_{b}$ each brought an equal contribution, because for any tree compatible with
$\phi_{a}$, there is a tree compatible with $\phi_{b}$ and vice--versa. Thus a single
variable $\phi$ was used for the two diagrams. 

The substitution of $\phi_{a}$ by $\phi_{b}$ in a pair of
diagrams does not always conserve the number of trees, so that the number of
independent variables for pairs of diagrams involving a $\phi$ cannot be reduced by a
factor 2. However, one may separate in $[\N_{2} \times \N_{3}]$, $[\N_{3} 
\times \N_{2}]$ and $\N_{3} \times
\N_{3}$ the trees for which the substitution is allowed from the others, like what we
did above regarding the change of arrows. 

It turns out that this is useful because only a reduced number of pairs of diagrams
misbehave under the change $\phi_{a} \leftrightarrow \phi_{b}$. Up to mirror 
symmetry, they are all given in Figure 3. 

\begin{figure}[htb]
\psfrag{L1}{$\La_1=$}
\psfrag{L2}{$\La_2=$}
\psfrag{L3}{$\La_3=$}
\psfrag{L4}{$\La_4=$}
\psfrag{P1}{$\Pi_1=$}
\psfrag{P2}{$\Pi_2=$}
\psfrag{P3}{$\Pi_3=$}
\psfrag{P4}{$\Pi_4=$}
\psfrag{Psi1}{$\Psi_1=$}
\psfrag{Psi2}{$\Psi_2=$}
\psfrag{Phi1}{$\Phi_1=$}
\psfrag{Phi2}{$\Phi_2=$}
\psfrag{Ome1}{$\Omega_1=$}
\psfrag{Ome2}{$\Omega_2=$}
\psfrag{L1p}{$\La_1^p=$}
\psfrag{L2p}{$\La_2^p=$}
\psfrag{L3p}{$\La_3^p=$}
\psfrag{L4p}{$\La_4^p=$}
\psfrag{P1p}{$\Pi_1^p=$}
\psfrag{P2p}{$\Pi_2^p=$}
\psfrag{P3p}{$\Pi_3^p=$}
\psfrag{P4p}{$\Pi_4^p=$}
\psfrag{Psi1p}{$\Psi_1^p=$}
\psfrag{Psi2p}{$\Psi_2^p=$}
\psfrag{Phi1p}{$\Phi_1^p=$}
\psfrag{Phi2p}{$\Phi_2^p=$}
\psfrag{Ome1p}{$\Omega_1^p=$}
\psfrag{Ome2p}{$\Omega_2^p=$}
\center{\mbox{\includegraphics[scale=.33]{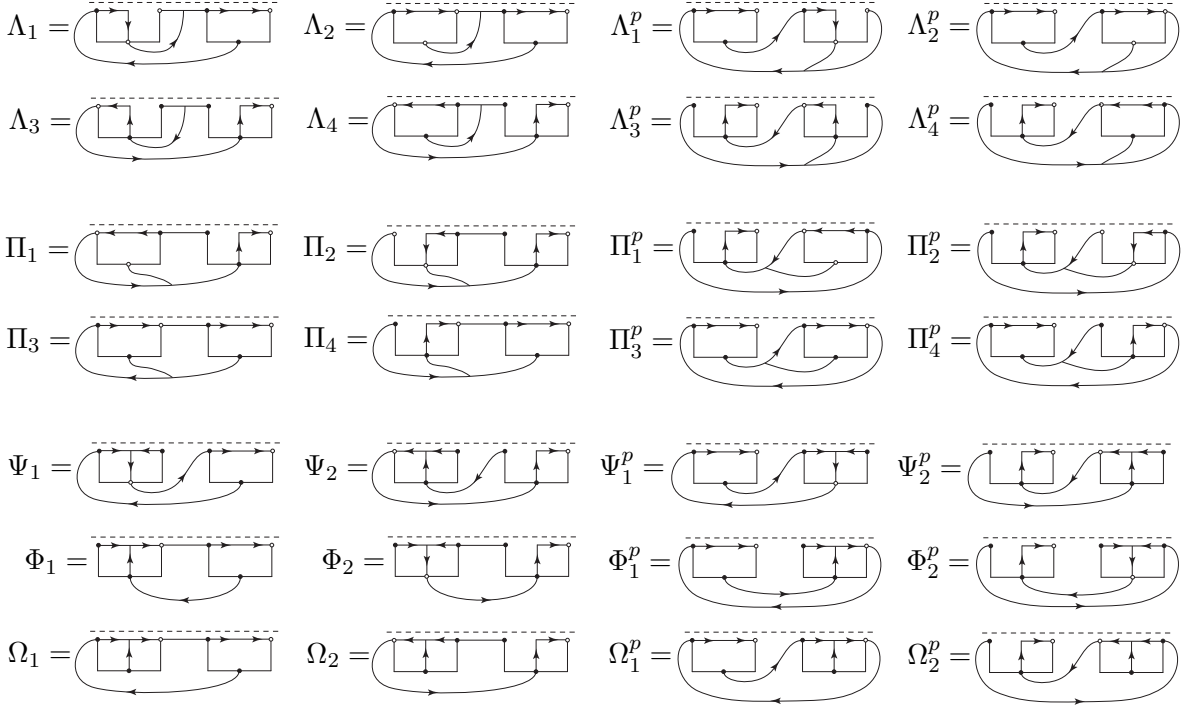}}}
\caption{\small Non local diagrams representing spanning trees which have an 
anomalous behaviour under the substitution of $\phi_{a}$ by $\phi_{b}$, or of 
$\phi_{b}$ by $\phi_{a}$. The superscript $p$ indicates that the two diagrams at the
reference sites have been permuted. All mirror diagrams must be added.}
\end{figure}

For instance the two diagrams $\Lambda_{1}$ and $\Lambda_{2}$ are pairs
$[\beta_2\phi_{a}]$ and $[\beta_{1}\phi_{a}]$, contained in the set $[\N_{2} \times
\N_{3}]$. The change $\phi_{a} \rightarrow \phi_{b}$ in $\Lambda_{1}$ requires a    
change of direction in the path going from the southern neighbour of $j_{0}$ through
$i_{0}$ and back to $j_{0}$, which is not possible. The pair $\Pi_{1}$ and $\Pi_{2}$
corresponds to the diagrams $[\tilde\beta_{1}\phi_{b}]$ and $[\tilde\beta_{2} 
\phi_{b}]$, also in $[\N_{2} \times \N_{3}]$. Their permuted versions belong to 
$[\N_{3} \times \N_{2}]$, while all the other diagrams in Figure 3 are in $\N_{3} 
\times \N_{3}$. The diagrams whose labels differ only by the numerical subscript
contribute equally, so $\Lambda_{1} = \Lambda_{2}$ but $\Lambda_{1} \neq 
\tilde\Lambda_{1}$.

If one denotes by curly brackets the sets of trees which are closed under the change 
$\phi_{a} \leftrightarrow \phi_{b}$, one can write
\bea
[\N_{2} \times \N_{3}] \egal \{[\N_{2} \times \N_{3}]\} + \Lambda_{1} + 
\Lambda_2 + \Pi_{1} + \Pi_2 + \tilde\Lambda^{p}_{1} + \tilde\Lambda^{p}_2 + 
\tilde\Pi^{p}_{1} + \tilde\Pi^{p}_2,\\
\left[\N_{3} \times \N_{2}\right] \egal \{[\N_{3} \times \N_{2}]\} + 
\Lambda^{p}_{1} +  \Lambda^{p}_2 + \Pi^{p}_{1} + \Pi^{p}_2+ 
\tilde\Lambda_{1} + \tilde\Lambda_2 + \tilde\Pi_{1} + \tilde\Pi_2,\\
\N_{3} \times \N_{3} \egal \{\N_{3} \times \N_{3}\} + \Lambda_{3} + 
\Lambda_4 + \Pi_{3} + \Pi_4 + \Phi_{1} + \Phi_{2} + \Psi_{1} + \Psi_{2} 
+ \Omega_{1} + \Omega_{2} \nonumber\\
&& \hspace{2.5cm} +\  {\rm mirrors\ and\ permuted}.
\eea

The diagrams $\Lambda_{1}+\Lambda_{2}$, $\Pi_{1}+\Pi_{2}$ and the permuted tilded 
versions contribute equally to $P_{23}$ and $P_{33}$, since they are in $[\N_{2} \times
\N_{3}]$. The diagrams $\Lambda^{p}_{1} + \Lambda^{p}_{2}$, $\Pi^{p}_{1} + 
\Pi^{p}_{2}$ (and the permuted tilded versions) contribute equally to $P_{32}$ and
$P_{33}$, whereas all the others contribute to $P_{33}$ only. Thus the expressions for the
2--site probabilities become
\bea
P_{22} \egal P_{12} + P_{21} - P_{11} + {[\N_{2} \times \N_{2}] \over 4\N} ,\\
P_{23} \egal P_{13} + P_{22} - P_{12} + {\{[\N_{2} \times \N_{3}]\} \over 2\N} 
+ {1 \over 2\N} [2A_{1} + 2B_{1} + 2C_{1} + 2C_{4} + 2\tilde A_2 + 
2\tilde B_2 + 2\tilde C_2 + 2\tilde C_5\nonumber\\
&&  +\  \Lambda_{1} + \Lambda_2 + \Pi_1 + \Pi_{2} 
+ \tilde\Lambda^{p}_{1} + \tilde\Lambda^{p}_2 + 
\tilde\Pi^{p}_{1} + \tilde\Pi^{p}_2],\\
P_{33} \egal P_{23} + P_{32} - P_{22} + {\{\N_{3} \times \N_{3}\} \over \N} 
\nonumber\\
&& +\  {1 \over \N} [A_{3} + B_{3} + C_{3} + C_{6} - A_{1} - A_{2} - B_{1} - 
B_{2} - C_{1} - C_{2} - C_{4} - C_{5} + {\rm mirrors}]\nonumber\\
&&  +\ {1 \over \N} [\Lambda_{3} + \Lambda_4 + \Pi_{3} + \Pi_4 +
 \Psi_{1} +  \Psi_{2} + \Phi_{1} + \Phi_{2}  + \Omega_{1} + \Omega_{2}
+ {\rm mirrors\ and\ permuted}].
\eea

The variables entering these expressions can be determined from the
same linear system as above, expressed in terms of the new variables. For instance, the
first equation becomes

\newcommand\figC{\parbox{115mm}{
\psfrag{Paa}{$Q_{1,1}\,=$}
\psfrag{e}{$=$}
\psfrag{p}{$+$}
\psfrag{betbet}{$[\beta\beta]$}
\psfrag{phiphi}{$\{\phi\phi\}$}
\psfrag{betaphi}{$\{[\beta\phi]\}$}
\psfrag{phibeta}{$\{[\phi\beta]\}$}
\psfrag{L}{$\Lambda_2$}
\psfrag{P}{$\Pi_3$}
\psfrag{Lp}{$\Lambda_2^p$}
\psfrag{Pp}{$\Pi_3^p$}
\psfrag{C}{$C_3$}
\includegraphics[scale=.33]{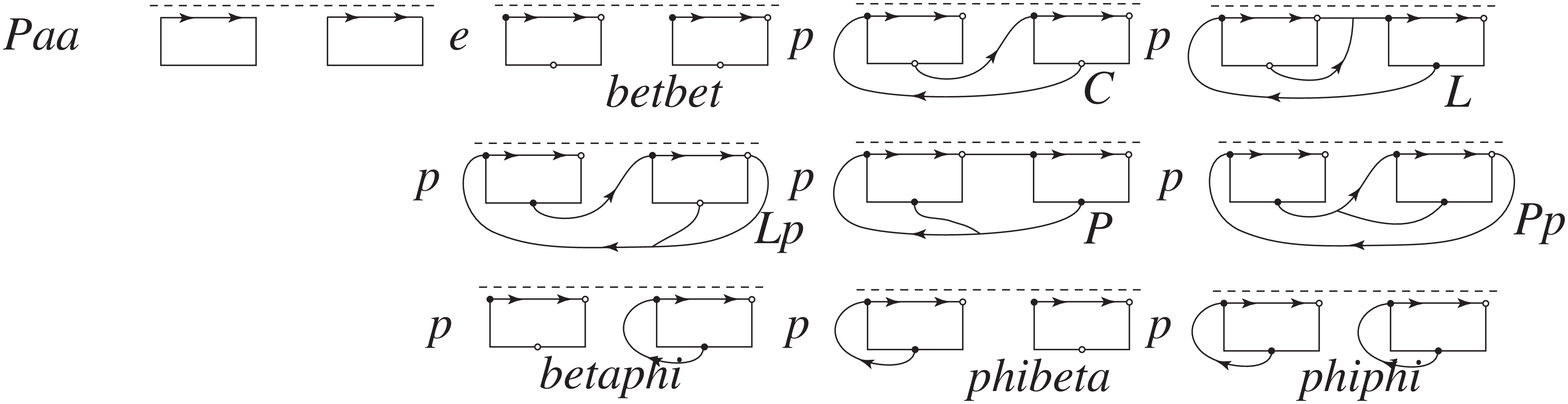}}}
\be
\figC
\ee

The number of equations is the same, but we have fewer variables. The new sets
$\{[\N_2 \times \N_3]\}$, $\{[\N_3 \times \N_2]\}$ and $\{\N_3 \times \N_3\}$
have respectively 15, 15 and 25 variables, to which 20 extra variables are
added for the diagrams in Figure 3 and their mirror images. So there are 88 variables,
constrained by 81 linear equations, of which 73 only are linearly independent. The
system is still underdetermined but unexpectedly allows to determine enough variables to
compute the probabilities without further work. 

First one can show, by suitably combining the independent equations, that 
$[A + B + C + \tilde C]/\N$  is equal to a combination of local diagrams which turns out
to be subdominant, of order $t^{3}$ for both an open or a closed boundary (order 
$m^{-6}$ at the critical point, where $m$ is the distance between $i_{0}$ and $j_{0}$). As
the four quantities are positive by construction, it means that each of them is at least of
order $t^{3}$, and can be neglected. Thus one relation determines four variables. Once
these four variables are eliminated, one is left with a system of 72 independent equations
for 84 variables.
 
Being independent, the 72 equations allow to determine 72 combinations of variables.
The point is that one can choose these 72 combinations in such a way that the
probabilities can be fully expressed in terms of them only, thereby making the knowledge
of the other 12 combinations useless. Alternatively, one may choose to solve the linear
system for 72 variables, which then become functions of the remaining 12. When inserted in 
the probabilities, all dependences in the 12 unknowns drop out completely. The set of
the 12 variables that the system cannot determine is not unique, but a possible choice is
$\{\Psi, \Psi^{p}, \Lambda^{p}, \Pi^{p}, \Phi^{p}, \Omega^{p} + {\rm mirrors}\}$.

The counting of variables and equations is different when there are no insertions at
other places than $i_{0}$ and $j_{0}$ since the mirrored equations are redundant. One
finds that the linear system is again not invertible, but is nonetheless sufficient to
compute all 2--site boundary probabilities. They have been computed in the massive
model to the dominant order $t^{2}$, which yields the universal terms. 

For an open boundary, we found that none of the diagrams in Figure 3 contribute to the
dominant order, being at least of order $t^{3}$. The probabilities $P_{22},P_{23}$ and
$P_{33}$ have the same form $t^{2} [K_{0}''(m\sqrt{t}) - K_{0}(m\sqrt{t})]^{2}$ 
at dominant order, and  only differ by their normalizations. These have been checked to be
in agreement with the identifications obtained in Section 4. 

The case of a closed boundary is a bit more complicated. In this case the diagrams of
Figure 3 contribute to order $t^{2}$ (as we have seen above, none of the diagrams of Figure
2 contribute, irrespective of the boundary condition), and the probabilities read
\def\ts{\textstyle}
\bea
P_{22}  \egal \ts  t^{2}\Big(-\frac{4}{\pi^{4}}K_{0}(m\sqrt{t})^{2}-
\left(\frac{48}{\pi^{4}}-\frac{12}{\pi^{3}}\right)K_{0}'(m\sqrt{t})^{2}+
\frac{2}{\pi^{3}}K_{0}(m\sqrt{t})K_{0}''(m\sqrt{t}) \nonumber\\
&& \ts \qquad\qquad-\left(\frac{144}{\pi^{4}}-\frac{72}{\pi^{3}}+
\frac{37}{4\pi^{2}}\right)K_{0}''(m\sqrt{t})^{2}+\left(\frac{12}{\pi^{3}}-
\frac{3}{\pi^{2}}\right)K_{0}'(m\sqrt{t})K_{0}'''(m\sqrt{t})\Big) +
\ldots\\
P_{23}  \egal \ts  t^{2}\Big(-\frac{1}{2\pi^{3}}K_{0}(m\sqrt{t})^{2}+
\left(\frac{8}{\pi^{4}}-\frac{3}{\pi^{3}}+\frac{3}{4\pi^{2}}\right)
K_{0}'(m\sqrt{t})^{2}-\left(\frac{1}{\pi^{3}}-\frac{1}{8\pi^{2}}\right)
K_{0}(m\sqrt{t})K_{0}''(m\sqrt{t}) \nonumber\\
&&\ts \qquad\qquad +\left(\frac{48}{\pi^{4}}-\frac{12}{\pi^{3}}+
\frac{1}{4\pi^{2}}\right)K_{0}''(m\sqrt{t})^{2}-\left(\frac{8}{\pi^{3}}-
\frac{3}{2\pi^{2}}\right)K_{0}'(m\sqrt{t})K_{0}'''(m\sqrt{t})\Big) +
\ldots\\
P_{33}  \egal \ts  t^{2}\Big(-\frac{1}{16\pi^{2}}K_{0}(m\sqrt{t})^{2}
+\frac{2}{\pi^{3}}K_{0}'(m\sqrt{t})^{2}-\frac{1}{4\pi^{2}}
K_{0}(m\sqrt{t})K_{0}''(m\sqrt{t}) \nonumber\\
&&\ts \qquad\qquad-\left(\frac{16}{\pi^{4}}+\frac{1}{4\pi^{2}}
\right)K_{0}''(m\sqrt{t})^{2}+\frac{4}{\pi^{3}}K_{0}'(m\sqrt{t})
K_{0}'''(m\sqrt{t})\Big) + \ldots
\eea
Again they are in full agreement with the fields found in Section 4.

We have also computed a few 3--site probabilities, when one of the insertion is a height 1
or a supercritical height. Then the same system as above can be used, the only difference
is that the Laplacian has to be decorated by a local defect matrix, and only affects
the calculation of the local diagrams. We have found for instance the connected probability
$P_{212}$ to have a height 1 and two heights 2 on a closed boundary, all separated by
large distances, at the critical point (the expressions for off--critical 3--site
probabilities are too long),
\bea
P_{212,\rm conn} \egal \frac{2}{\pi ^{3}}\Big({3\over4}-{2\over\pi}\Big)
\left(\frac{1}{m_{12}m_{13}^{\,2}m_{23}^{\,3}}+\frac{1}{m_{12}^{\, 3}
m_{13}^{\,2}m_{23}}+\frac{1}{\pi ^{2}}\frac{(6\pi -24)^{2}}{m_{12}^{\,2}
m_{13}^{\,2}m_{23}^{\, 2}}\right.\nonumber\\
&&  + \left.\frac{6\pi -24}{\pi }\left[\frac{1}{m_{12}m_{13}^{\, 3}
m_{23}^{\,2}}+\frac{1}{m_{12}^{\, 2}m_{13}^{\, 3}m_{23}}+
\frac{1}{m_{12}^{\,3}m_{13}m_{23}^{\, 2}}+\frac{1}{m_{12}^{\, 2}
m_{13}m_{23}^{\,3}}\right]\right)+\ldots 
\eea
where $m_{ij}$ is the distance between the $i^{\mathrm{th}}$ and $j^{\mathrm{th}}$ 
site. The connected probability is equal to $P_{212,\rm conn} = P_{212} - P_{2}P_{12} -
P_{22}P_{1} - P_{21}P_{2} + 2P_{2}^{2}P_{1}$. The previous formula for $P_{212}$ 
is equivalent to that found by Jeng \cite{jeng1}, but allows for a more direct
comparison with the field theoretic result, as the various terms correspond to 
specific Wick contractions.

\section*{Acknowledgments}

We would like to thank Monwhea Jeng for instructive discussions and comparisons of our
respective methods. They led us to realize that our early calculations had overlooked
some diagrams, which however had no effect on the end results.

\end{document}